\documentstyle [12pt] {article}

\catcode`@=12
\begin{document}
\thispagestyle{empty}
\begin{flushright}
UCRHEP-T115\\
MAD/PH/792\\
September 1993\
\end{flushright}
\vspace{0.5in}
\begin{center}
{\large \bf New Supersymmetric Electroweak Gauge Model\\
with Natural Baryon-Number Conservation\\}
\vspace{1.5in}
{\bf Ernest Ma\\}
\vspace {0.1in}
{\sl Department of Physics, University of California, Riverside,
California 92521\\}
{and\\}
{\sl Department of Physics, University of Wisconsin, Madison, Wisconsin
53706\\}
\vspace {1.5in}
\end{center}
\begin{abstract}\
Baryon number (B) is automatically conserved in the standard SU(2) $\times$
U(1)
electroweak gauge model, but not in its supersymmetric extension where it must
be
imposed as an extra condition.  This undesirable feature is avoided in the
conventional left-right supersymmetric extension, but then flavor-changing
neutral
currents (FCNC) are not naturally suppressed.  A new version is proposed with
automatic B conservation as well as FCNC suppression (except for $u$ quarks
through scalar exchange.)

\end{abstract}
\newpage
\baselineskip 24pt

In the standard SU(2) $\times$ U(1) electroweak gauge model, because of its
assumed
particle content and the requirement of gauge invariance, it is not possible to
write down renormalizable interaction terms which violate the conservation of
baryon number (B).  This is a very desirable feature of the model since
B conservation
is experimentally known to be very well respected.  On the other hand, if the
model
is extended to include supersymmetry, then such terms are in general allowed.
Consider the quark and lepton superfields.  In a notation where only the left
chiral projections are counted, they transform under SU(3) $\times$ SU(2)
$\times$
U(1) as follows:  $Q \equiv (u,d) \sim (3,2,1/6)$, $ u^c \sim
(\overline 3, 1, -2/3)$,
$d^c \sim (\overline 3, 1, 1/3)$,  $L \equiv (\nu,e) \sim (1,2,-1/2)$,
and $e^c \sim
(1,1,1)$, where the family index has been suppressed.  In addition,
there must be two
Higgs superfields $\Phi_{1,2}$
transforming as $(1,2, \mp 1/2)$ respectively.  The desirable allowed terms in
the
superpotential are then $\Phi_1 Q d^c$, $\Phi_2 Q u^c$, and $\Phi_1 L e^c$,
which
supply the quarks and leptons with masses as the neutral scalar components of
$\Phi_{1,2}$ acquire nonzero vacuum expectation values $v_{1,2}$.
However, the terms
$u^c d^c d^c$, $L Q d^c$, and $L L e^c$ are also allowed {\it a priori}
and they violate
the conservation of B as well as that of lepton number (L).  Note also
that $\Phi_1$
and $L$ are indistinguishable by their transformations alone.

To obtain a realistic model, the usual solution is to impose B and L
conservation
as an extra condition.  For example, if a discrete Z$_2$ symmetry is assumed
such
that $\Phi_{1,2}$ are even, but all other superfields are odd, then
only the desirable
terms are retained.  Since both B and L are now conserved by assumption, a
quantity
$\rm R \equiv (-1)^{2j+3B+L}$ can be used to distinguish particles
($\rm R = +1$) from
superparticles ($\rm R = -1$).  This has become such common practice
that its origin
as an imposed condition is often accepted without much attention.  However, as
long
as there is a need to impose some extra symmetry on the supersymmetric standard
model, the choice of that symmetry becomes somewhat arbitrary.  For example, if
$L$ and $e^c$ are even instead of odd in the above, then the $u^c d^c d^c$
terms
are still forbidden, but the $L Q d^c$ and $L L e^c$ terms are allowed.  Hence
B
is conserved, but L is not.  Such models have been studied in the past\cite{1}
and limits on the various possible Yukawa couplings have been obtained, usually
under certain restrictive assumptions.  It is also possible to refine the
imposed
condition so that instead of preserving all three lepton numbers, {\it i.e.}
L$^e$,
L$^\mu$, and L$^\tau$, only one linear combination (say L$^e -$ L$^\mu$) or two
(say L$^e$ and L$^\mu$) are conserved.\cite{2}  Finally, it may be assumed that
$Q$, $u^c$, and $d^c$ are even instead of odd, then L is conserved but B is
not.
This may also lead to some rather interesting phenomenology.\cite{3}

Given that the terms $u^c d^c d^c$, $L Q d^c$, and $L L e^c$ are problematic in
the
supersymmetric standard model, it is natural to ask whether such terms can be
avoided automatically in a larger theory.  The answer is yes if we simply take
the conventional left-right supersymmetric extension of the standard model.
The
gauge group is now $\rm SU(3) \times SU(2) \times SU(2) \times U(1)$.  The
quarks
and leptons are
\begin{eqnarray}
Q \equiv (u,d) \sim (3,2,1,1/6), ~&~&~ Q^c \equiv (d^c,u^c) \sim (\overline 3,
1,
2, -1/6),  \\ L \equiv (\nu,e) \sim (1,2,1,-1/2), ~&~&~ L^c \equiv
(e^c,N^c) \sim (1,1,2,1/2).
\end{eqnarray}
Hence terms involving three such superfields are not possible in the
superpotential
because of gauge invariance, and the automatic conservation of B and L is
achieved.
On the other hand, realistic quark and lepton masses require the existence of
{\underline {two}} bidoublet Higgs superfields $\Phi_{1,2} \sim (1,2,2,0)$ with
{\underline {four}} nonzero vacuum expectation values, and flavor-changing
neutral currents (FCNC) are not naturally suppressed.  It is perhaps for this
reason that no detailed analysis of the conventional left-right supersymmetric
model exists in the literature.

An alternative left-right supersymmetric model with natural flavor conservation
can
be constructed\cite{4} if we add an extra quark $h$ of charge $-1/3$ so that we
may
repalce $d^c$ by $h^c$ in $Q^c$, leaving $h$ and $d^c$ as singlets.  In that
case,
each scalar vacuum expectation value is matched to a different quark type,
and the natural suppression of FCNC is assured.  However, an extra
discrete symmetry
is still required to distinguish $h$ from $d$ and it is this same symmetry
which
forbids the undesirable terms $Q Q h$ and $Q^c Q^c d^c$.

Consider instead now the addition of an extra quark $x$ of charge 2/3 with no
extra
symmetry. Let
\begin{equation}
x \sim (3,1,1,2/3), ~~~ x^c \sim (\overline 3, 1, 1, -2/3),
\end{equation}
then it is again not possible to have terms in the superpotential involving
three
quark or lepton superfields, and the desirable feature of automatic B and L
conservation is achieved.  The Higgs superfields are
\begin{equation}
\Phi_{12} \equiv \left( \begin{array} {c@{\quad}c} \overline {\phi_1^0} &
\phi_2^+
\\ -\phi_1^- & \phi_2^0 \end{array} \right) \sim (1,2,2,0),
\end{equation}
and
\begin{equation}
\Phi_3 \equiv (\overline {\phi_3^0}, -\phi_3^-) \sim (1,1,2,-1/2), ~~~
\Phi_4 \equiv (\phi_4^+,\phi_4^0) \sim (1,1,2,1/2).
\end{equation}
Note that both $\Phi_3$ and $\Phi_4$ are singelts under the standard SU(2)
$\times$ U(1) and the above structure is anomaly-free.  The allowed Yukawa
terms
are then
\begin{equation}
\Phi_{12} Q Q^c = \overline {\phi_1^0} d d^c + \phi_1^- u d^c + \phi_2^0 u u^c
-
\phi_2^+ d u^c,
\end{equation}
and
\begin{equation}
\Phi_3 x Q^c = \overline {\phi_3^0} x u^c + \phi_3^- x d^c.
\end{equation}
Since $x$ and $x^c$ are singlets, there is also an allowed gauge-invariant
mass term $xx^c$.  Let $\langle \phi^0_{1,2,3} \rangle = v_{1,2,3}$, then
the 6 $\times$ 6 mass matrix linking ($u,x$) with ($u^c,x^c$) is given by
\begin{equation}
{\cal M}_{ux} = \left[ \begin{array} {c@{\quad}c} v_2 v_1^{-1} {\cal M}_d &
0  \\ {\cal M}_3  & {\cal M}_x \end{array} \right],
\end{equation}
where the 3 $\times$ 3 matrices ${\cal M}_d$ and ${\cal M}_x$ can be defined
to be diagonal, and ${\cal M}_3$ is proportional to $v_3$.
Note that $\Phi_4$ is not involved at all and it will be assumed here that
$\langle \phi_4^0 \rangle = 0$.
The mixing of $u$ and $x$ is determined by the matrix
\begin{equation}
{\cal M}_{ux} {\cal M}_{ux}^\dagger = \left[ \begin{array} {c@{\quad}c}
v_2^2 v_1^{-2} {\cal M}_d {\cal M}_d^\dagger & v_2 v_1^{-1} {\cal M}_d
{\cal M}_3^\dagger \\ v_2 v_1^{-1} {\cal M}_3 {\cal M}_d^\dagger &
{\cal M}_3 {\cal M}_3^\dagger + {\cal M}_x {\cal M}_x^\dagger \end {array}
\right],
\end{equation}
and since ${\cal M}_d$ should be very much smaller than ${\cal M}_3$ and
${\cal M}_x$, it is clear that $u-x$ mixing is very small and can be safely
neglected.  On the other hand, the mass matrix for the $u$ quarks is
given by
\begin{equation}
{\cal M}_u {\cal M}_u^\dagger = v_2^2 v_1^{-2} {\cal M}_d \left[ 1 -
{\cal M}_3^\dagger ({\cal M}_3 {\cal M}_3^\dagger + {\cal M}_x
{\cal M}_x^\dagger )^{-1} {\cal M}_3 \right] {\cal M}_d^\dagger,
\end{equation}
which is in general nondiagonal and
can easily be phenomenologically correct even if $u-x$ mixing is
very small.    The mixing of $u^c$ and $x^c$ is determined
by the matrix ${\cal M}_{ux}^\dagger {\cal M}_{ux}$ and can be quite large
because ${\cal M}_3$ and ${\cal M}_x$ should be comparable in magnitude for a
realistic ${\cal M}_u$ to be obtained.  However, because $u^c$ and $x^c$
transform identically under the standard SU(2) $\times$ U(1),
this mixing is observable at the electroweak energy scale only through
scalar interactions.  Nevertheless, this means that FCNC for the $u$
quarks exist and they should be searched for in processes such as
$D^0 - \overline {D^0}$ mixing.

In the leptonic sector, the analogous Yukawa term to Eq. (6) is
\begin{equation}
\Phi_{12} L L^c = \overline {\phi_1^0} e e^c + \phi_1^- \nu e^c + \phi_2^0 \nu
N^c
- \phi_2^+ e N^c,
\end{equation}
which by itself would imply that $\nu$ and $N^c$ should form a Dirac neutrino
with mass equal to $v_2 v_1^{-1} m_e$.  That is of course unrealistic.  The
usual
solution of this problem in the conventional left-right model is to introduce a
complex Higgs triplet which provides a large Majorana mass for $N^c$, thus
breaking L in the process.  Here it is simpler just to add a singlet superfield
$N$ (with zero vacuum expectation value) and the Yukawa term analogous to Eq.
(7) is
\begin{equation}
\Phi_3 L^c N = \overline {\phi_3^0} N^c N + \phi_3^- e^c N.
\end{equation}
Noting that a gauge-invariant
Majorana mass for $N$ is allowed, the 9 $\times$ 9 Majorana mass
matrix spanning $\nu, N^c$, and $N$ is then given by
\begin{equation}
{\cal M} = \left[ \begin{array} {c@{\quad}c@{\quad}c} 0 & v_2 v_1^{-1}
{\cal M}_\ell
& 0 \\ v_2 v_1^{-1} {\cal M}_\ell & 0 & {\cal M}'_3 \\ 0 &
{\cal M}'_3 & {\cal M}_N \end{array} \right],
\end{equation}
where the 3 $\times$ 3 mass matrices ${\cal M}_\ell$ and ${\cal M}_N$
can be defined
to be diagonal, and ${\cal M}'_3$ is proportional to $v_3$.
Under the reasonable assumption $v_2 v_1^{-1} {\cal M}_\ell \ll
{\cal M}'_3 \ll {\cal M}_N$,
the neutrino mass matrix is then
of the double-see-saw form:
\begin{equation}
{\cal M}_\nu = v_2^2 v_1^{-2} {\cal M}_\ell {\cal M}_3^{\prime -1} {\cal M}_N
{\cal M}_3^{\prime -1} {\cal M}_\ell.
\end{equation}
Naturally small neutrino masses as well as mixing among the three neutrinos are
now possible and a realistic lepton sector is obtained.

Recall that in the supersymmetric standard model, one of the Higgs superfields,
namely $\Phi_1$, is identical to the doublet lepton superfields $L$ in its
transformations.  Now there is no confusion with $L$, but instead we have
$\Phi_4$ and $L^c$ which transform in the same way.  This means that the terms
$\Phi_3 \Phi_4$,
$\Phi_3 L^c$, $\Phi_3 \Phi_4 N$, and $\Phi_{12} \Phi_4 L$ are allowed.
However, we can always redefine $\Phi_4$ and $L^c$ so that the $\Phi_3 L^c$
term is absent.  Now since $\langle \phi_3^0 \rangle = v_3$ but neither
$\Phi_4$ nor $L^c$ has a vacuum expectation value, the 5 $\times$ 5 mass
matrix linking ($\ell, \psi_3^-, \tilde w_R^-$) with ($\ell^c, \psi_4^+,
\tilde w_R^+$), where $\tilde w_R^\pm$ are the charged $\rm SU(2)_R$
gauginos and $\psi_{3,4}^\mp$ are the corresponding higgsinos,
is of the form
\begin{equation}
{\cal M} = \left[ \begin{array} {c@{\quad}c@{\quad}c} {\cal M}_\ell &
{\cal M}_{12} & 0 \\ 0 & \mu_{34} & M_{W_R} \sqrt 2 \\ 0 & 0 & M_R
\end{array} \right],
\end{equation}
where ${\cal M}_{12}$ is a 3 $\times$ 1 matrix coming from the $\Phi_{12}
\Phi_4 L$ term, and $M_R$ is an allowed $\rm SU(2)_R$ gauge-invariant
term which softly breaks the supersymmetry.  Since $\mu_{34}$ is expected
to be very much greater than ${\cal M}_{12}$ in magnitude,
the mixing of $\ell$ with $\psi_3^-$ and $\tilde w_R^-$ can easily
be made negligible and
lepton universality at the electroweak energy scale is maintained.
However, the diagonalization of the above mass matrix will involve
some small mixing of the leptons with the charginos so that
flavor-changing neutral currents will be present, but only at a
very much reduced level.

Because of the choice of gauge group and particle content, the proposed
supersymmetric model has the following desirable features.  (1) The
conservation of baryon number (B) is automatic.  (2) The Higgs sector is
minimal, consisting of just one bidoublet $\Phi_{12}$ and two
${\rm SU(2)_R \times U(1)}$ doublets $\Phi_{3,4}$.  (3) Realistic mass
matrices for both quarks and leptons are obtained.  (4) Flavor-changing
neutral currents (FCNC) are naturally suppressed at the electroweak
energy scale, except for $u$ quarks through scalar exchange.
(5) The conservation
of lepton number (L) is violated only through the singlet $N$ and the
fact that both $L^c$ and $\Phi_4$ have identical transformations
and the term $\Phi_3 \Phi_4$ is unavoidable.  The
only significant effect of this at low energies is the
appearance of small Majorana masses for the neutrinos.
Rare decays violating lepton number are possible, but they are
suppressed by inverse powers of some higher mass scale.  For example,
the rate for $\mu \rightarrow eee$ through Z exchange is proportional
to $\mu_{34}^{-4}$.

At or below the electroweak energy scale of $10^2$ GeV, the proposed
left-right supersymmetric model has the same particle content of the
minimal supersymmetric standard model (MSSM).  Their gauge interactions
are of course identical but their
Yukawa and scalar interactions do have important differences.
For example, the physical neutral Higgs
boson corresponding to the breaking of the standard SU(2) $\times$ U(1),
{\it i.e.} $\rm Re\phi_1^0 cos \beta + Re\phi_2^0 sin \beta$, where $\tan
\beta \equiv v_2/v_1$, couples to $\overline u u$ with strength
proportional to $v_2 v_1^{-1} m_d ~\times$ a factor due to $u^c - x^c$
mixing in this model, instead of the well-known $m_u$ in
the MSSM.  The detailed phenomenology of this and other effects will be
given elsewhere.

At the electroweak energy scale, only the two doublets contained in
$\Phi_{12}$ are possibly relevant to the Higgs potential $V$ of this
model. Let $\Phi_{1,2} = (\phi_{1,2}^+, \phi_{1,2}^0)$, then $V$ is
of the well-known form
\begin{eqnarray}
V &=& \mu_1^2 \Phi_1^\dagger \Phi_1 + \mu_2^2 \Phi_2^\dagger \Phi_2 +
\mu_{12}^2 (\Phi_1^\dagger \Phi_2 + \Phi_2^\dagger \Phi_1) \nonumber \\
&+& {1 \over 2} \lambda_1 (\Phi_1^\dagger \Phi_1)^2 + {1 \over 2}
\lambda_2 (\Phi_2^\dagger \Phi_2)^2 + \lambda_3 (\Phi_1^\dagger \Phi_1)
(\Phi_2^\dagger \Phi_2) + \lambda_4 (\Phi_1^\dagger \Phi_2) (\Phi_2^\dagger
\Phi_1).
\end{eqnarray}
Using the supersymmetric constraints of the SU(2) $\times$ SU(2) $\times$
U(1) gauge group and taking into account\cite{5} the cubic interactions of
$\Phi_{1,2}$ with Re$\phi_3^0$ after its spontaneous breakdown to
the standard SU(2) $\times$ U(1),
the above quartic scalar couplings are determined to be
\begin{equation}
\lambda_1 = \lambda_2 = {1 \over 4} (g_1^2 + g_2^2),~~\lambda_3 = -{1 \over 4}
g_1^2 + {3 \over 4} g_2^2,~~\lambda_4 = - g_2^2,
\end{equation}
where $g_1$ and $g_2$ are the U(1) and SU(2) gauge couplings of the standard
model respectively.  Note that $\lambda_{3,4}$ are not the same as in the
MSSM where they are equal to $(-g_1^2+g_2^2)/4$ and $-g_2^2/2$ respectively.
We now have the tree-level sum rule
\begin{equation}
m_{H^\pm}^2 = m_A^2 + 2 M_W^2,
\end{equation}
where $m_A$ is the mass of the pseudoscalar boson, instead of the well-known
$m_{H^\pm}^2 = m_A^2 + M_W^2$ in the MSSM.  The other tree-level sum rule
\begin{equation}
m_{H_1^0}^2 + m_{H_2^0}^2 = m_A^2 + M_Z^2
\end{equation}
remains the same, but radiative corrections due to the Yukawa couplings
of the Higgs bosons will be different in this model.
Note also that even though there
are no cubic terms in the superpotential, $V$ does not reduce to that of the
MSSM as in previous examples\cite{6}.  The reason is that in this model,
the soft supersymmetry-breaking terms of the Higgs potential at the
SU(2) $\times$ SU(2) $\times$
U(1) level must be chosen to maintain the hierarchy $v_{1,2} \ll v_3$
in such a way that the two scales cannot be entirely separated.
As a result,
\begin{equation}
m_A^2 = - {M_{W_R}^2 \over {\cos 2 \beta}}
\end{equation}
and the Higgs
sector reduces at the electroweak energy scale to that of a single
physical particle as in the standard model.  Hence $\cos 2 \beta
< 0$, or equivalently $\tan \beta > 1$, is required in this model.
Details will be given elsewhere.

Lepton number is violated here either because the singlet $N$ has a
Majorana mass or because the terms $\Phi_{12} L L^c$, $\Phi_{12} \Phi_4
L$, and $\Phi_3 \Phi_4$ must coexist.  Its effect is very much
suppressed for ordinary processes.  One important consequence for
future experimental searches for supersymmetry is that there is no
longer a ``lightest supersymmetric particle.''  The photino for
example will mix with the neutrinos and will decay.  However, if the
mixing turns out to be very small and the photino mass not too heavy,
it may still be stable within the detector in a typical high-energy
physics experiment.

Finally, it should be mentioned that if left-right exchange symmetry is
desired, we can easily add two $\rm SU(2)_L$ Higgs doublets $\Phi_{5,6}
\sim (1,2,1,\pm 1/2)$ without affecting anything essential in this paper
as long as $\Phi_6$, which transforms identically to $L$, does not have
any vacuum expectation value.  A conundrum of the MSSM is that $\Phi_1$
and $L$ transform identically and yet $\Phi_1$ must have a vacuum
expectation value, but $L$ must not.  Here the only superfields which need
to have vacuum
expectation values are $\Phi_{12}$ and $\Phi_3$, both of which are
in representations different from the leptons.

\vspace{0.3in}
\begin{center} {ACKNOWLEDGEMENT}
\end{center}

The author thanks V. Barger for discussions and for reading the
manuscript. This work was supported in part by the U. S. Department
of Energy under contracts No. DE-AT03-87ER40327 and No. DE-AC02-76ER00881,
by the Texas National Laboratory Research Commission, and by
the University of Wisconsin Research Committee
with funds granted by the Wisconsin Alumni Research Foundation.

\newpage
\bibliographystyle{unsrt}

\end{document}